\documentclass[lettersize,journal]{IEEEtran}
\usepackage{amsmath,amsfonts}
\usepackage{algorithm}
\usepackage{algorithmicx}  
\usepackage{array}
\usepackage[caption=false,font=normalsize,labelfont=sf,textfont=sf]{subfig}
\usepackage{textcomp}
\usepackage{stfloats}
\usepackage{url}
\usepackage{verbatim}
\usepackage{graphicx}
\usepackage{cite}
\usepackage{algorithm}
\usepackage{algpseudocode}
\usepackage{bm}
\usepackage{algorithmicx}  
\usepackage{algpseudocode}
\usepackage{setspace}
\usepackage{authblk} 
\usepackage{cite}
\usepackage{pifont}
\usepackage{xcolor}

\begin{document}

\title{Prototyping and Experimental Results for 
Environment-Aware Millimeter Wave   Beam
Alignment via
Channel Knowledge Map }

\author{{Zhuoyin Dai, Di Wu, Zhenjun Dong, Kun Li, Dingyang Ding, \\Sihan Wang, and
Yong Zeng,~\IEEEmembership{Senior Member, IEEE} 
  }
\thanks{
   

The authors are with the National Mobile Communications Research Laboratory,
Southeast University, Nanjing 210096, China. Y. Zeng is also with the 
Purple Mountain Laboratories, Nanjing 211111, 
China (e-mail: \{zhuoyin\_dai, studywudi, zhenjun\_dong, 220205571, 
220200693, turquoise,  yong\_zeng\}@seu.edu.cn). 
(\emph{Corresponding author: Yong Zeng.})

}}

\markboth{ }%
{Shell \MakeLowercase{\textit{et al.}}: A Sample Article Using IEEEtran.cls for IEEE Journals}


\maketitle

\begin{abstract}
        Channel knowledge map (CKM), which aims to directly reflect the intrinsic 
        channel properties of
        the local wireless environment, is a novel technique for achieving environment-aware 
        communication. In this paper, to alleviate the large training overhead in 
        millimeter wave (mmWave) beam alignment, an environment-aware and training-free 
        beam alignment prototype is established based on a typical CKM, termed beam index map (BIM).
        To this end, a general CKM construction method is first  presented, and 
        an indoor BIM is constructed offline to learn the candidate transmit and 
        receive beam index pairs for each grid in the experimental area.
        Furthermore, based on the location information of the receiver 
        (or the dynamic obstacles) from the ultra-wide band (UWB) positioning system, 
        the established BIM is used to achieve training-free beam alignment 
        by directly providing the beam indexes for the transmitter and receiver.
        Three typical scenarios are considered in the experiment, 
        including quasi-static environment with line-of-sight (LoS) link, quasi-static environment 
        without LoS link and dynamic environment.
        Besides, the receiver orientation measured from the gyroscope is also 
        used to help CKM  predict more accurate beam indexes.
        The experiment results show that compared with the benchmark location-based beam 
        alignment strategy, the CKM-based beam 
        alignment strategy can achieve much higher received power, 
        which is close to that achieved by exhaustive beam search,  
        but with significantly reduced training overhead.

\end{abstract}

\begin{IEEEkeywords}
        Channel knowledge map, environment-aware communication, training-free beam alignment,  millimeter wave.
\end{IEEEkeywords}

\section{Introduction}
Millimeter wave (mmWave) massive multiple-input multiple-output (MIMO) is an effective technology 
to meet the ever-increasing  capacity requirement
for the  fifth-generation (5G) and beyond  mobile communication networks \cite{mumtaz2016mmwave,heath2016overview,rappaport2013millimeter}.
On the one hand, the  
mmWave band at 26.5-300 GHz has abundant   spectrum, 
which can effectively alleviate the spectrum crunch  issue. 
On the other hand, the short
wavelengths at mmWave frequencies  allow  more antennas
to be compactly packed at the transmitter 
and receiver to compensate for the path loss. 
Although large antenna arrays bring considerable beamforming  gains, the overall system 
overhead  also increases significantly, not 
only in terms of  the   hardware cost, but also   the 
high overhead to practically  achieve  the large-dimensional beamforming gain. 
Some cost-effective
techniques have been proposed for
mmWave massive MIMO communication.  For example, 
hybrid analog/digital beamforming has been extensively studied to reduce 
the cost of radio frequency (RF) chains \cite{ zhang2005variable,
Venkateswaran2010analog,Ayach2014spatially,alkhateeb2014channel},
where   signal processing is divided into the digital domain 
and analog domain.  
Deploying low-precision digital-to-analog converters (ADC) is 
also a possible way to reduce hardware costs and power consumption in mmWave systems \cite{alkhateeb2014mimo}.
Lens antenna array is another cost-effective mmWave communication technology, 
which utilizes electromagnetic lenses to separate signals 
from different directions without requiring sophisticated signal 
processing \cite{zeng2016millimeter,gao2017reliable,zeng2017cost,yang2021communication}.

It is important to note that the radio propagation environment has 
a significant effect  on 
the performance of communication systems. 
In practice, 
it is typically necessary to acquire real-time channel state information (CSI) 
to completely realize the performance gain of highly directional beamforming. 
To this end, one major approach 
is to use training-based CSI estimation for beamforming design 
\cite{lee2016channel,venugopal2017channel,alkhateeb2014channel}. 
However,   
for cost-effective implementations such as hybrid beamforming, 
the increasing size of antennas  may significantly  increase 
the overhead for CSI acquisition, since 
the channel measured in the digital baseband is entangled with 
the analog beamforming for training.
An alternative beamforming method for 
mmWave massive MIMO is training-based beam sweeping 
\cite{giordani2018tutorial,heng2021six,kim2014elements,kim2014fast}. 
Instead of estimating the MIMO channels for beam selection, 
this method sequentially sweeps the beamforming vectors in the predefined codebook 
to find  the optimal transmit/receive beamforming pair.  
The exhaustive beam sweeping approach 
evaluates each beamforming pair by a number of metrics, such as 
the received power. Therefore, as the number of transmit/receive antennas 
and the size of the codebook continue to increase, 
the beam sweeping method will also incur prohibitive overhead \cite{giordani2018tutorial}.

It is worth remarking that the aforementioned beam alignment methods
always require obtaining accurate and real-time CSI, 
which is extremely difficult for mmWave systems with limited RF links. 
Besides, these existing methods do not exploit prior local environment 
information that is specific to the locations of the receiver and transmitter,
and this is a waste of valuable wireless environmental characteristics.
Recently,  a novel concept called \emph{ channel knowledge map} (CKM) has
 been proposed in \cite{CKM} 
in order to utilize  prior local information of 
the actual wireless environment for the design and optimization 
of future communication systems. 
Specifically,  CKM
is a site-specific database, tagged with the locations of the 
transmitters and/or receivers, that provides location-specific
channel knowledge useful to enhance environment-awareness 
and facilitate or even obviate sophisticated real-time CSI
acquisition \cite{CKM,Zeng2023ATO}.
Unlike
physical environment maps \cite{seidel1994site}, 
CKM can directly reflect the intrinsic channel characteristics 
that can be utilized for communication 
without the need for additional complex computations  such as ray tracing. Meanwhile, 
compared to conventional radio environment maps (REMs) that mainly concern spectrum usage 
information \cite{yilmaz2013radio,bi2019engineering}, CKM focuses  on the 
location-specific wireless channel knowledge  that is independent of the
transmitter/receiver activities. 
Theoretical studies on CKM construction and utilization have received increasing 
attention recently. For example, the expectation-maximization algorithm is utilized in \cite{li2022channel} 
to  construct  CKM by combining both expert knowledge about channel model 
and local measurement data, thus reducing the workload of 
the actual channel measurements. An environment-aware  beam alignment 
scheme with CKM is proposed in \cite{wu2022environment}, which
utilizes a specific type of CKM, namely beam index map (BIM), to significantly reduce the 
computational cost and training overhead for real-time beamforming. 
The idea of using fingerprint maps like BIM  to achieve beam alignment is 
similarly presented in \cite{va2018inverse}.
CKM-based environment-aware beamforming approach was also studied for 
the IRS-aided communication  \cite{ding2021environment}.

Although theoretical studies have demonstrated the 
great potential  of CKM-based environment-aware  communication, 
to the best of our knowledge, no prototyping experimental studies 
have been conducted to  verify the effectiveness of CKM in practical communication systems. 
To fill this gap, in this paper, we develop a prototyping experiment for CKM-based 
environment-aware communication for mmWave beam alignment. The main contributions of this work are summarized as follows:
\begin{itemize}
\item 
First, a general offline CKM construction method is   presented. 
The method selects a typical type of CKM, called BIM, as an example, 
and explores the intrinsic relationship between channel information, 
location information and environment information. Specifically, the target area 
is divided into equidistant grids in accordance with the required precision. 
For each grid, the coordinates of its center are measured as the grid coordinates. 
An exhaustive  beam sweeping is utilized to find the transmit and 
receive beam index pairs of each grid to complete the offline BIM construction. 
The proposed generic offline CKM construction method is applicable to 
both quasi-static and dynamic scenarios with line-of-sight (LoS) and 
non-line-of-sight (NLoS) links.

\item 
A prototype of the CKM-based environment-aware mmWave communication system is completed 
through the integration of program design and hardware construction. 
The prototype consists of ultra-wide band (UWB) positioning module, 
 mmWave phased array (mmPSA),  universal software radio peripheral (USRP), 
and  gyroscope. During the communication process, the prototype obtains the 
stored channel knowledge such as LoS and NLoS links from the already built CKM 
based on the receiver's location to achieve training-free real-time mmWave 
beam alignment. The prototype also uses the orientation of the receiver 
from the gyroscope to refine and predict more accurate transmit and 
receive beam index pairings.

\item 
Finally, the constructed CKM-based environment-aware mmWave communication prototype is 
experimentally verified. The performance of the CKM-based communication prototype 
is tested practically in a variety of  
quasi-static and dynamic scenarios. 
Compared with  the exhaustive   beam sweeping benchmark strategy, 
the CKM-based mmWave beam alignment strategy reduces the 
overhead from scanning 4096 beam index pairs to training-free.
Furthermore, compared to the location-based beam alignment, 
the proposed CKM-based strategy obtains significantly greater received power 
in various scenarios. This demonstrates the 
feasibility and effectiveness of CKM for environment-aware 
mmWave communication.

\end{itemize}

The rest of this paper is organized as follows. 
Section II  introduces the typical scenario of environment-aware communication 
that utilizes the CKM-based beam alignment strategy, and describes the 
theoretical basis of CKM.
The hardware equipment  and program design of the prototype are 
introduced in Section III. The construction of the CKM-based environment-aware
mmWave communication prototype  is specified in Section IV. 
Meanwhile,  Section IV also shows the experimental results
of the CKM-based   beam alignment, 
and the comparison  with the location-based benchmark is presented. 
  Finally, the conclusion is drawn in Section V.

\emph{Notations:} Scalars are denoted by italic letters.
Vectors and matrices are denoted by boldface lower- and upper-case letters,
respectively. $\mathbb{C}^{N\times M}$ denotes the
space of $N\times M$-dimensional complex-valued matrix.
 $\mathbf I_N$ denotes an $N \times N$ identity matrix.
For a matrix $\mathbf{A}$, its
transpose, conjugate, Hermitian transpose, and determinant  are respectively
denoted as
$\mathbf{A}^{\mathrm{T}}$,$\mathbf{A}^{*}$, $\mathbf{A}^{\mathrm{H}}$ and $| \mathbf{A}|$. 
For a vector $\mathbf{a}$, $[\mathbf{a}]_{i}$ denotes its $i$-th element.
In addition,
$\mathcal{CN}(\mu,\sigma^{2})$  denotes the 
circularly symmetric complex  Gaussian (CSCG)
distribution with mean $\mu$ and variance $\sigma^{2}$.

\section{Typical Scenario  and Basic Concepts}

\subsection{Communication Scenario}
\begin{figure}[!t]
        \centering{\includegraphics[width=1\columnwidth]{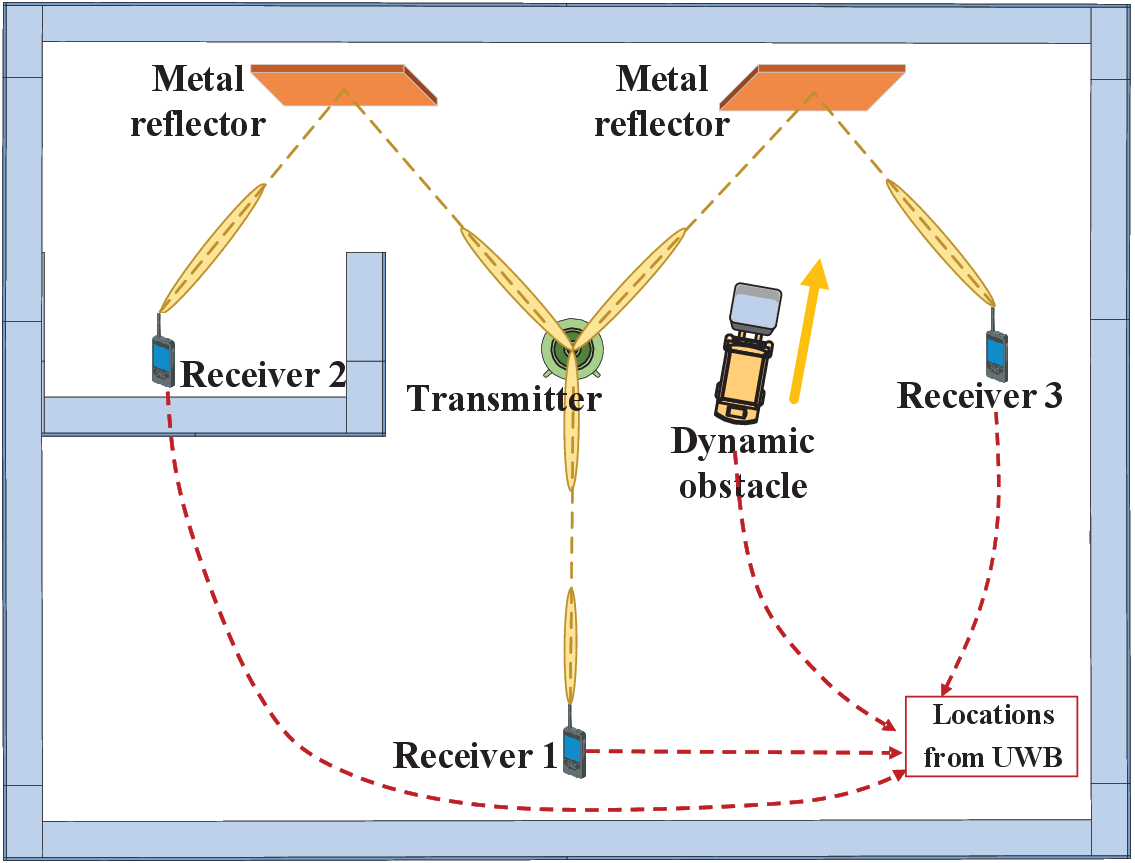}}%
        \caption{A typical scenario for CKM-based environment-aware communication.}
        \label{fig:framebasic}
\end{figure}

As shown in Fig. \ref{fig:framebasic}, we consider a typical mmWave massive 
MIMO communication scenario. A mmWave transmitter is located in the center of 
the considered area, and needs  to serve three receivers.
Considering the complexity of the communication environment, it is of 
great significance to select a suitable   strategy to achieve 
beam alignment between the mmWave transmitter and receiver, thus enhancing the 
mmWave communication performance. Specifically, we first consider  receiver 1 in 
Fig. \ref{fig:framebasic}. The LoS link with good channel conditions is available 
between the transmitter and receiver 1,
and can be utilized
for beam alignment and communication. Several  strategies can be applied to obtain the optimal transmit 
and receive beam pairs in this case, such as  exhaustive beam sweeping, 
location-based  beam calculation, and  the proposed CKM-based beam alignment.
Second, we consider the case of communication with  receiver 2.
Note that   receiver 2 is located in a room with thick walls, which block  the 
LoS link between the transmitter and   receiver 2.
The blocking of the LoS link renders the location-based beam calculation invalid, 
since it will lead to a beam pair
directed to the LoS link. On the other hand, exhaustive beam sweeping and the 
 CKM-based beam alignment 
can still find the strong reflected link provided by the metal reflector, thus enabling  communication
between the transmitter and   receiver 2. 
For   receiver 3, although there is no static blockage of the LoS link, 
a moving vehicle constitutes a dynamic obstruction in the communication environment. 
When the dynamic obstacle moves between   receiver 3 
and the transmitter, the LoS link is obstructed.
In this situation, the location-based beam calculation can only provide the 
transmit and receive beam pair corresponding to  the LoS link, resulting in  poor performance. On the 
other hand, although   exhaustive beam sweeping can switch the beam pair from the LoS link
to the reflected link,   it requires prohibitive delay and communication resources 
for sweeping all the 
possible beam pairs. Different from these two benchmark strategies, the CKM-based beam alignment 
switches the transmit and receive beam pair to the reflected link swiftly
according to the location of the dynamic obstacles
with the assistance of UWB positioning system to maintain the communication with the receiver 3.
 
\subsection{Basic concept of CKM}
\begin{figure}[htbp]
	\centering{\includegraphics[width=.4\textwidth]{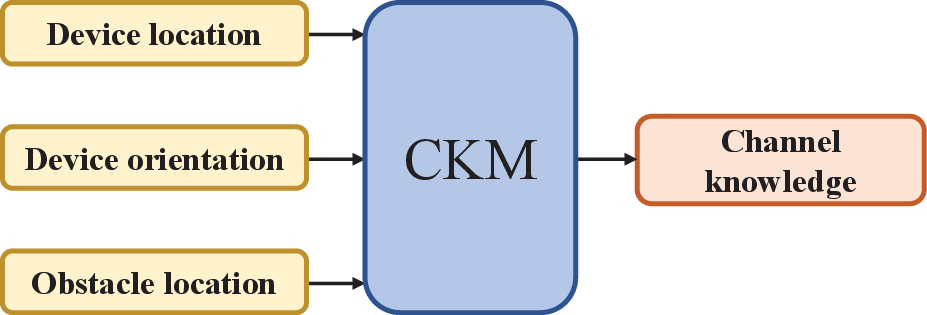}}  	
	\caption{An illustration of CKM for environment-aware communication.} \label{CKM}  
\end{figure}

CKM aims to provide the channel knowledge (such as path loss, 
angle-of-arrival (AoA)/angle-of-departure (AoD), beam indexes) based on 
location and orientation  information of mobile devices and major obstacles, as illustrated in Fig. \ref{CKM}.
Based on the acquired channel knowledge, the channel/beam can be 
reconstructed/selected directly, which is 
termed \emph{CKM-based training-free communication}
\cite{CKM,wu2021environment,ding2021environment,wu2022environment}.

In this paper, we choose one specific type of CKM, named BIM, 
as an example in the experiment. Different from other types of CKM that store some 
specific channel knowledge such as gains, AOAs/AODs, etc., BIM directly stores 
several candidate indexes of transmit and receive beam pairs for each location of 
interest. With BIM, BS and UE can obtain candidate beam indexes based on their 
locations and implement refined beam sweeping within them, instead of exhaustively 
searching all the beam pairs.
To construct the BIM, we divide the region of interest  $ \mathcal{Q} $ 
into $ P  $ small areas, denoted as  $ \mathbf{Q}=\{\mathcal{Q}_1,...,\mathcal{Q}_P\} $.
BIM aims to learn the indexes of the possible beam pairs for all 
potential receiver locations $ \mathcal{Q}_p $ within the coverage of 
interest $ \mathcal{Q}$, which can be expressed as 
\begin{equation}
	\text{BIM}: \quad \{\mathcal{Q}_p;(\hat{\mathcal{F}}_p,\hat{\mathcal{W}}_p)\}_{p=1}^{P}
	\label{BIMmapping}
\end{equation}
 where $ \hat{\mathcal{F}}_p $ and $ \hat{\mathcal{W}}_p $ are subset of 
 the transmit and receive beam codebooks $ \mathcal{F} $ and $ \mathcal{W} $, i.e., $ \hat{\mathcal{F}}_p\subset \mathcal{F},\hat{\mathcal{W}}_p\subset \mathcal{W}$, which have significantly reduced size $ |\hat{\mathcal{F}}_p|\ll |\mathcal{F}|$ and $|\hat{\mathcal{W}}_p|\ll |\mathcal{W}|$.

Based on the real-time location information $ \mathbf{q} $ from the localization system, 
the area of the UE can be determined and the corresponding possible beam index pairs 
can be obtained from BIM. 
In BIM-based training-free communication, the   beam index pair is selected for data transmission and reception, 
without additional real-time training.

\section{Prototyping Experiment Implementation}
\subsection{Prototype Architecture}

\begin{figure}
        \centering
         \includegraphics[width=1\columnwidth]{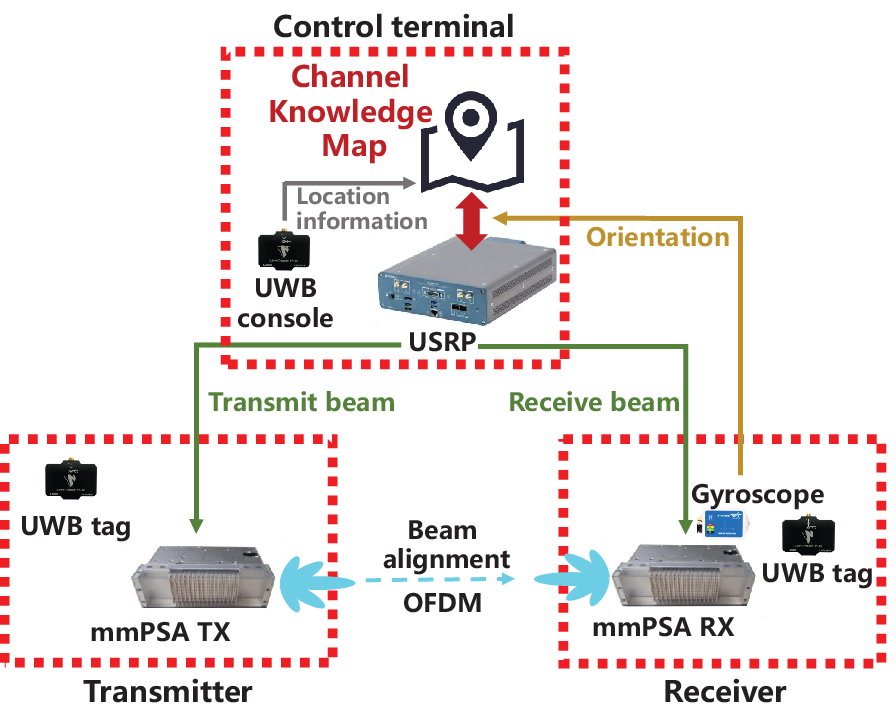}
        \caption{The logical system for CKM-based mmWave wireless communication.}
        \label{fig:frame_0}
\end{figure}
The logical architecture of the prototyping system for CKM-based beam alignment
is shown in Fig. \ref{fig:frame_0}, which is composed of the transmitter,
the receiver, and the controller for CKM-based beam alignment by obtaining the receiver location information from the UWB positioning system and the receiver orientation from the gyroscope.
The construction of CKM is completed offline in advance.
In the process of online real-time communication, the system first
obtains the location and orientation information  of the receiver from the UWB positioning system and gyroscope, respectively.
The controller determines in which grid the receiver is located according to the location information,
and obtains the transmit beam corresponding to
that grid directly  according to the CKM constructed offline. 
Furthermore, the receive beam index is obtained based on 
both the receiver's location and orientation information.
Finally, the transmit and receive beam indexes are forwarded to the transmitter and receiver, respectively,
and the transmitter and receiver then complete beam alignment for communication by using the corresponding beam.

\subsection{Hardware Equipment }
\begin{figure}
        \centering
         \includegraphics[width=1\columnwidth]{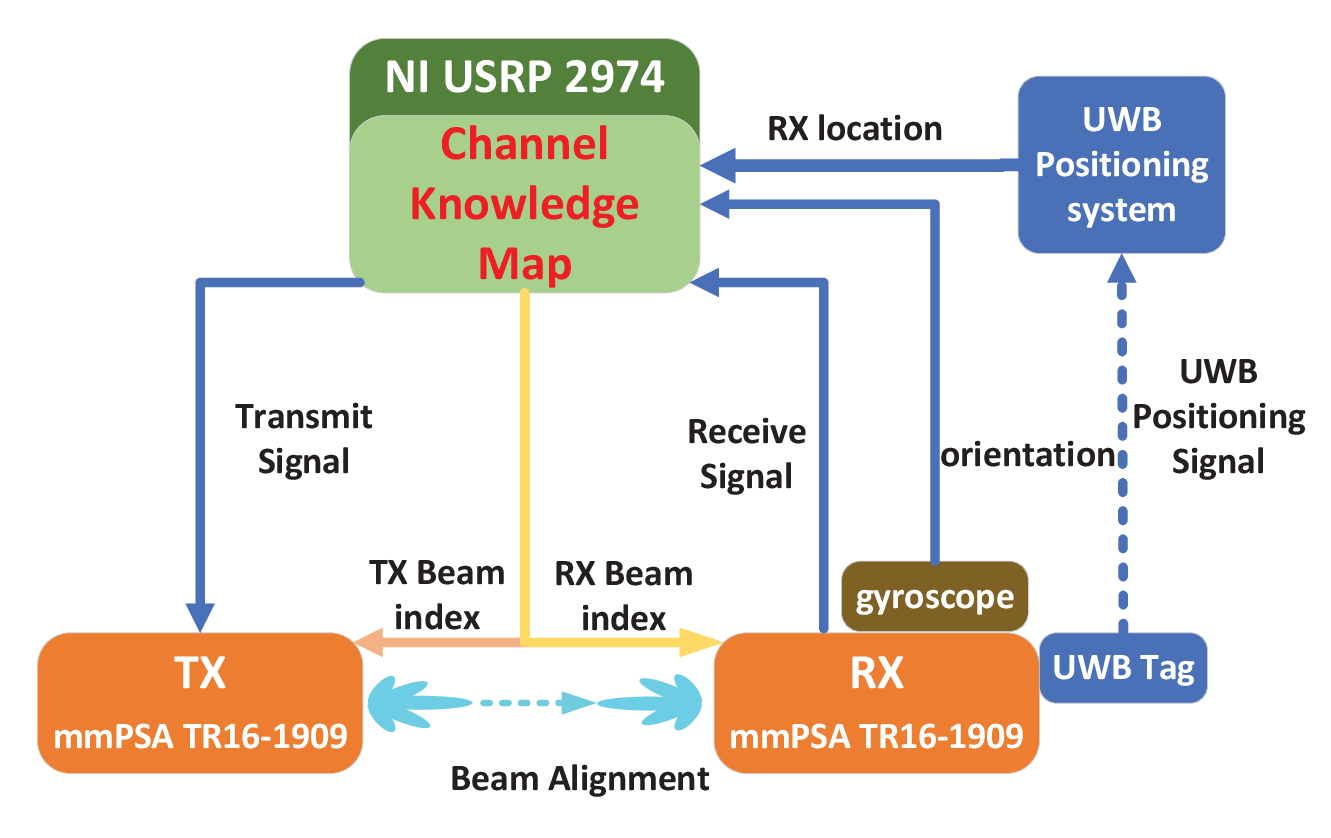}
        \caption{The hardware system for implementing the CKM-based mmWave wireless communication.}
        \label{fig:frame}
\end{figure}

As shown in Fig. \ref{fig:frame}, the hardware system is composed of four modules 
containing the NI USRP 2974 module for implementing the function of the controller, 
the mmPSA TR16-1909 module for mmWave phased antenna array, 
the UWB indoor positioning system, and the gyroscope module for obtaining the receiver's orientation.
\subsubsection{NI USRP 2974}
The NI USRP-2974 is a Software Defined Radio (SDR) stand-alone device built with
a Field-Programmable Gate Array (FPGA) and an onboard processor for rapidly prototyping high-performance
wireless communication systems. With a real time bandwidth of up to 160MHz, two daughter boards covering $10$MHz to $6$GHz are
utilized to perform analog up/down conversion. The baseband signal sampling/interpolation and digital up/down conversion are realized rapidly on the Xilinx Kintex-7 410T FPGA, while the Intel i7 onboard processor is responsible for waveform related operations like modulation/demodulation. Configured by the NI LabView Communication program, USRP-2974 can be operated as a transmitter to generate required signals flexibly and a receiver to process signals efficiently.

\subsubsection{mmPSA TR16-1909}
The mmPSA TR16-1909 is used as a phased antenna array  for  mmWave communication systems, 
which is a 16-element  phased array and provides vertical polarization service.
mmPSA TR16-1909 is cable-connected to USRP, to which the control frame is transmitted from the NI LabView Communications program to
realize exhaustive beam sweeping. There are two control frame transmission
methods, namely  the SPI interface via HDMI line and the RS232 interface via USB-RJ45 connection line.

The device supports time division duplex (TDD) communication,
 whose operating frequency band is 27-29 GHz. The phase array provides the
 fast and normal modes in phase control. In the fast mode,
 16 phase modules can be set in dependently, and the setting accuracy of
 each phase module is 5 bits. And in the normal mode, 16 phase modules are set
 simultaneously, and the setting accuracy  
 is increased by 1 bit compared to the fast mode.
The mmPSA TR16-1909 has a codebook with a scale of 64,
where the angular range is from $-56^{\circ}$ to $+56^{\circ}$ in the
horizontal direction.

\subsubsection{Nooploop UWB Indoor Positioning System}
The UWB indoor positioning system is used to measure the location
of the receiver and transmitter in our experiment.
The UWB technology  can obtain
Gbit/s data rate   by transmitting nanosecond or shorter pulses. 
Compared with the traditional narrowband system, the UWB system 
has superior anti-interference performance, a
high transmission rate, an extremely wide bandwidth, a large system capacity, low
transmission power, good confidentiality, etc.
Therefore, it has obvious advantages in short-range high-speed data transmission and 
high-precision indoor positioning.
The Nooploop UWB   positioning system used in the CKM-based beam alignment
prototype has a positioning update rate of 200Hz and a positioning accuracy of 10cm.

The UWB indoor positioning system is composed of the equipment layer, solution layer, and application layer,
mainly including UWB base station, UWB tag, Power Over Ethernet (POE) switch, positioning engine, LabView program platform, etc.
The main workflow is as follows:   UWB tags are attached to the transmitter and receiver,
sending the measurement signals to the base stations in the scenario through the UWB channel.
In this experiment, five positioning base stations are set up, 
which collect tag data and transmit it  to the solution engine 
in order to compute the UWB tag coordinates. After that,
the parsed location information data 
is   forwarded and uploaded to the LabView host computer and can be visualized in the LabView program platform,
so as to control the mmPSA and perform  CKM-based beam alignment based on the location information.

\subsubsection{Gyroscope Module}
The gyroscope module can efficiently obtain the  real-time motion attitude
with an attitude measurement accuracy of $0.2^{\circ}$ and extremely high stability, 
which adopts a high-performance microprocessor and advanced dynamic filtering algorithm.
 It can provide real-time data with an update rate of up to 200Hz, so as to meet the 
 needs of various high-precision applications and accomplish accurate motion capture and 
 attitude estimation.

\subsection{Program Design}
\begin{figure}
  \centering
   \includegraphics[width=1\columnwidth]{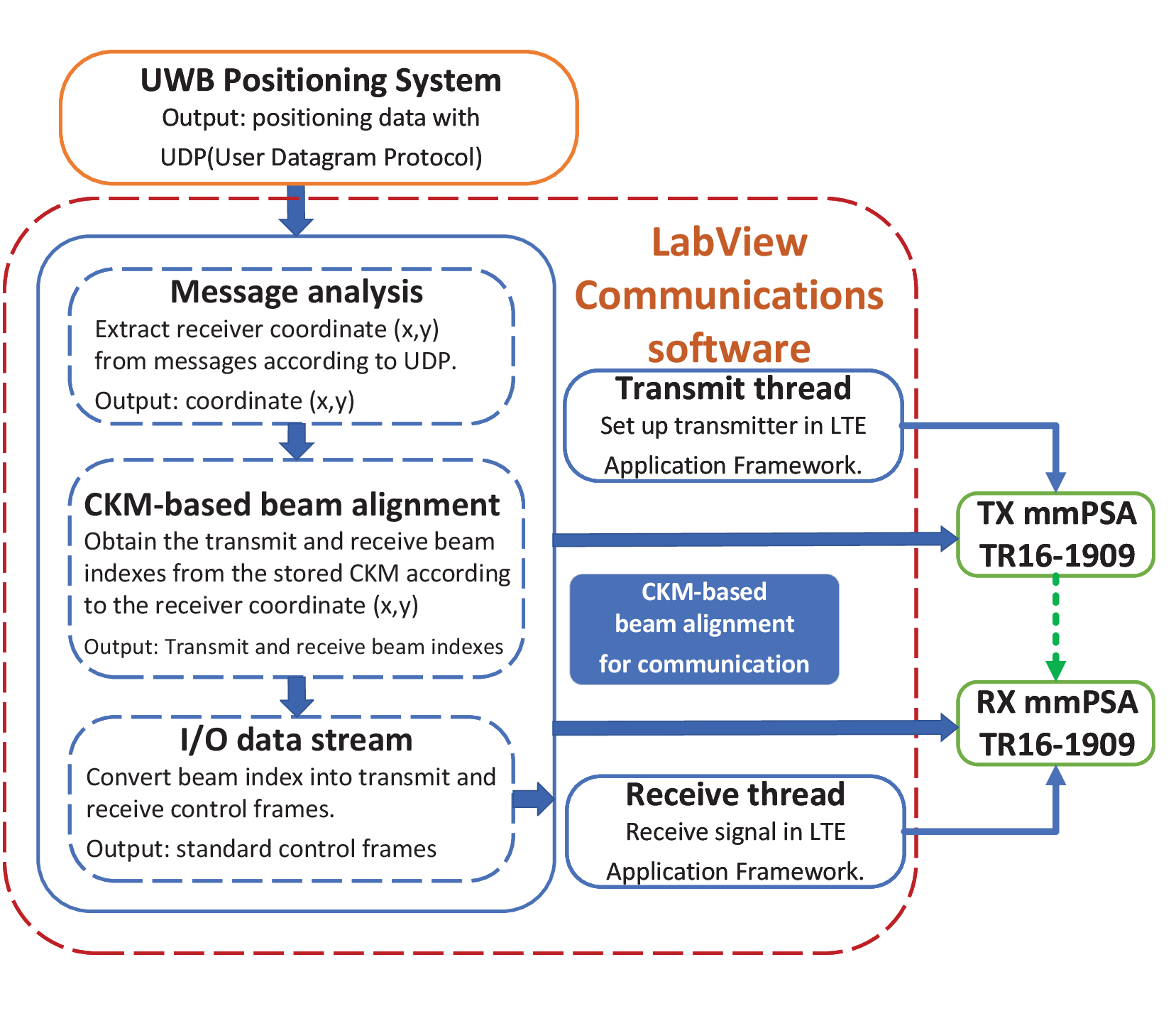}
  \caption{ The LabView communications program architecture for the CKM-based mmWave beam alignment prototype.}
  \label{fig:Software}
\end{figure}

The experiments are conducted by using the LabView Communications program installed on the USRP 2974.
The purpose of program design is to select the optimal beam index  pair from the codebook of mmPSA
to realize reliable real-time communication, when the location and orientation information of
the receiver changes dynamically. 
As shown in Fig. \ref{fig:Software}, the designed program consists of three 
parallel primary threads: the transmit thread, the receive thread, 
and the CKM-based beam alignment thread.

In the designed program, the transmitter is always in the state of transmitting signals, and the quality of the received signal is then determined by
observing the constellation diagram or the received power at the receiver. 
If the location or orientation information of the receiver changes significantly 
while the transmit and receive beam indexes remain unchanged, the received signal power can be observed to become lower and
the constellation diagram becomes worse. Therefore, it is of paramount importance to select the optimal beam index pair
in the CKM-based beam alignment thread to maintain high-quality communication.
In the CKM-based beam alignment thread, the location data packets transmitted by the UWB positioning system are  parsed to get the coordinate of the receiver first. Then, the transmit and receive beam indexes are obtained from the stored CKM according to the receiver's coordinate. Furthermore, the receive beam index is modified according to the orientation information of the receiver obtained from the gyroscope module.
Finally, the transmit and receive beam indexes are converted into 
standard control frame format,  which are sent to the transmit and receive 
mmPSA TR16-1909 via USB-RJ45 connection line, respectively.

\section{ Experimental Results}
Prototyping experiments are conducted to demonstrate the  
environment-aware communication system with CKM-based beam alignment strategy. 
The common 
parameters of the experiments are listed as follows: 
1) The selected mmPSAs are  fully activated 16-antenna linear arrays.
2) The data transmission frequency of these prototypes is set as 28GHz.
3) Multiple modulation modes including QPSK, 16QAM, and 64QAM are used.
In the experiments, the location-based beam alignment strategy is selected as 
the benchmark for comparison.
The scenarios involved in the experiments include quasi-static and dynamic scenarios, 
and communication through both LoS and reflective NLoS links are considered.

\subsection{Benchmark: Location-based Beam Alignment }

The location-based beam alignment strategy generally assumes that there are no obstacles 
in the physical environment between the transmitter and the receiver. Therefore, this strategy 
considers the LoS link as the dominant link and performs beam alignment based on 
the LoS link. In the LoS case, the classical geometry-based channel model with a 
fixed transmitter can be written as
\begin{equation}
	\mathbf{H}[t]=g(\mathbf{q}[t])
        =\sqrt{M_rM_t}\gamma[t]\mathbf{a}_r(\theta_{\mathrm{A}}[t])
        \mathbf{a}_t^H(\theta_{\mathrm{D}}[t]),
	\label{H_loc}
\end{equation}
where $\mathbf{q}[t]=[ {q}_x[t], {q}_y[t]]$ is the location of the receiver,
$ \gamma[t] $ is the complex gain of the LoS path, $ \theta_{\mathrm{A}}[t] $ 
and $ \theta_{\mathrm{D}}[t] $ are the AoA and AoD.
Besides, $ \mathbf{a}_t(.) $ and $ \mathbf{a}_r(.) $ represent the transmit and receive array response vectors, respectively.
In this case, the beam alignment problem is written as 
\begin{equation}
        \hat{\mathbf{f}}[t]=\mathop{\arg \max}_{\mathbf{f} \in \mathcal{F}} 
        |\mathbf{a}_t^H(\theta_{\mathrm{D}}[t])\mathbf{f}|^2,\\
\end{equation}
\begin{equation}
        \hat{\mathbf{w}}[t]=\mathop{\arg \max}_{\mathbf{w} \in \mathcal{W}} 
        |\mathbf{a}_r^H(\theta_{\mathrm{A}}[t])\mathbf{w}|^2,
        \label{fw_loc}
\end{equation}
where $\mathbf{f}$ and $\mathbf{w}$ are arbitrary beamforming vectors in
the transmit and receive beam codebooks $ \mathcal{F} $ and $ \mathcal{W} $, 
while $\hat{\mathbf{f}}$ and $\hat{\mathbf{w}}$ are the selected 
transmit and receive beamforming vectors, respectively.

\begin{figure}[!t]
        \centering
       \includegraphics[width=1\columnwidth]{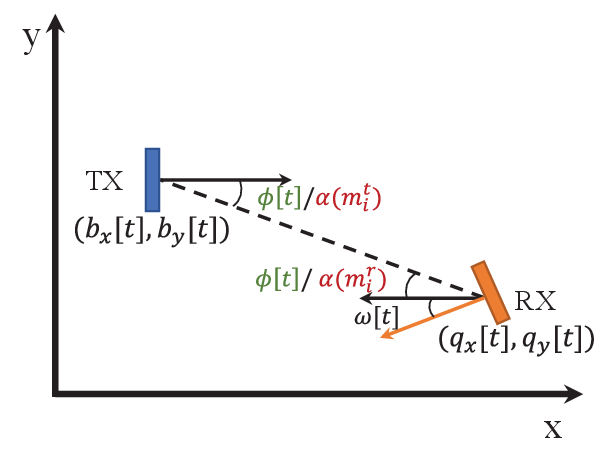}
        \caption{A geometric illustration of the location/CKM-based beam alignment
        in LoS scenario.}
        \label{fig:location}
\end{figure}

For the location-based beam alignment strategy, location information 
is the key to performing beam alignment.
This strategy mainly uses the receiver coordinate $[ {q}_x[t], {q}_y[t]]$ and the 
transmitter coordinate $[b_{x}[t],b_{y}[t]]$ 
to calculate the angle $\phi[t]$ with respect to the x-axis, and then selects the 
transmit and receive beam index pair from the mmWave phased array codebook 
based on $\phi[t]$.
According to the geometric relationship as shown in  Fig. \ref{fig:location}, 
the angle $\phi[t]$ between transmitter and receiver can be calculated as 
\begin{equation}\label{eq:phit}
        \phi[t]=\arctan\left(\frac{q_y[t]-b_y[t]}{q_x[t]-b_x[t]}\right).	
\end{equation}


\begin{figure}[!t]
        \centering
        \subfloat[]{\includegraphics[width=1\columnwidth]{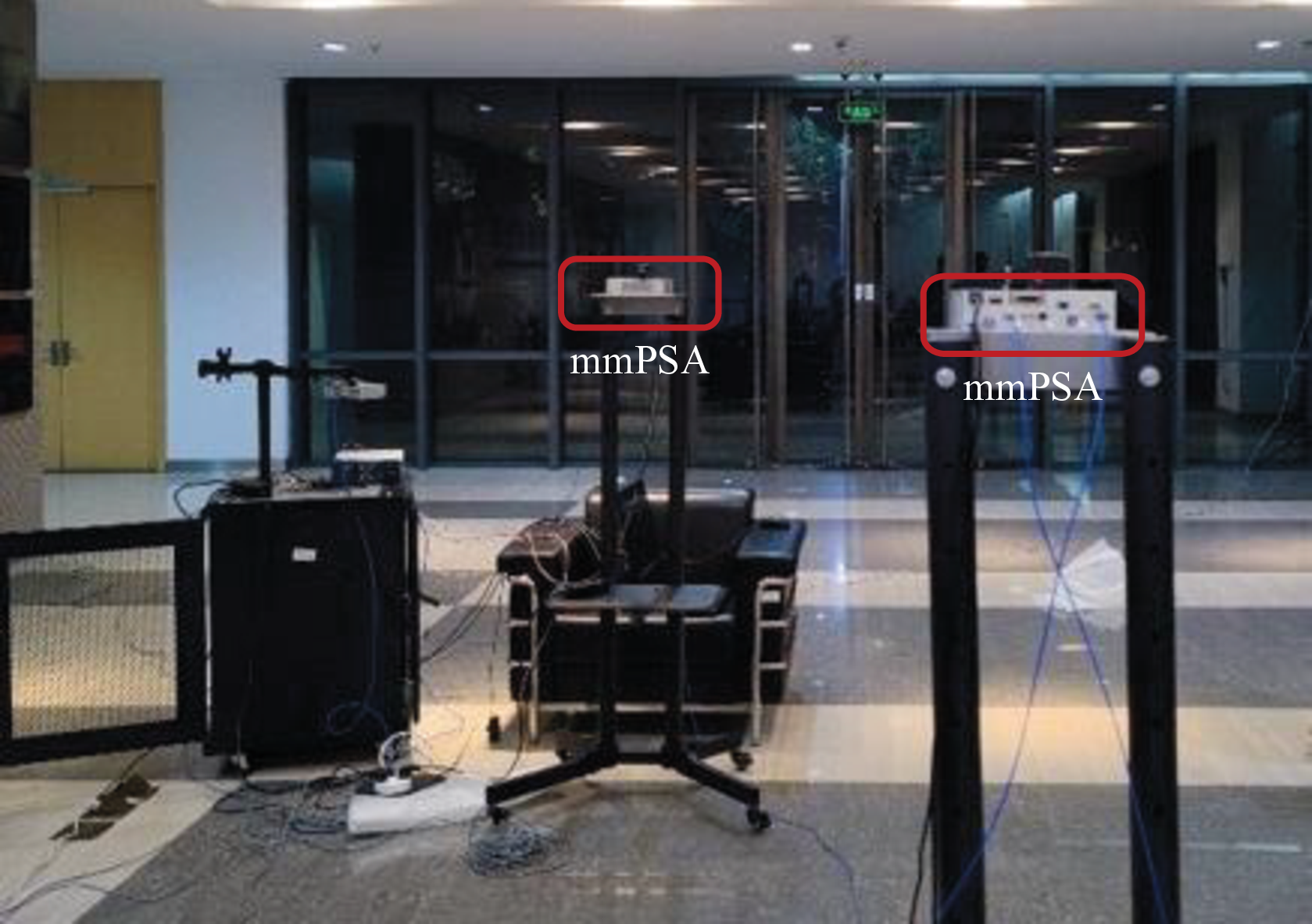}%
        \label{fig:1a}}
        \hfil
        \subfloat[]{\includegraphics[width=1\columnwidth]{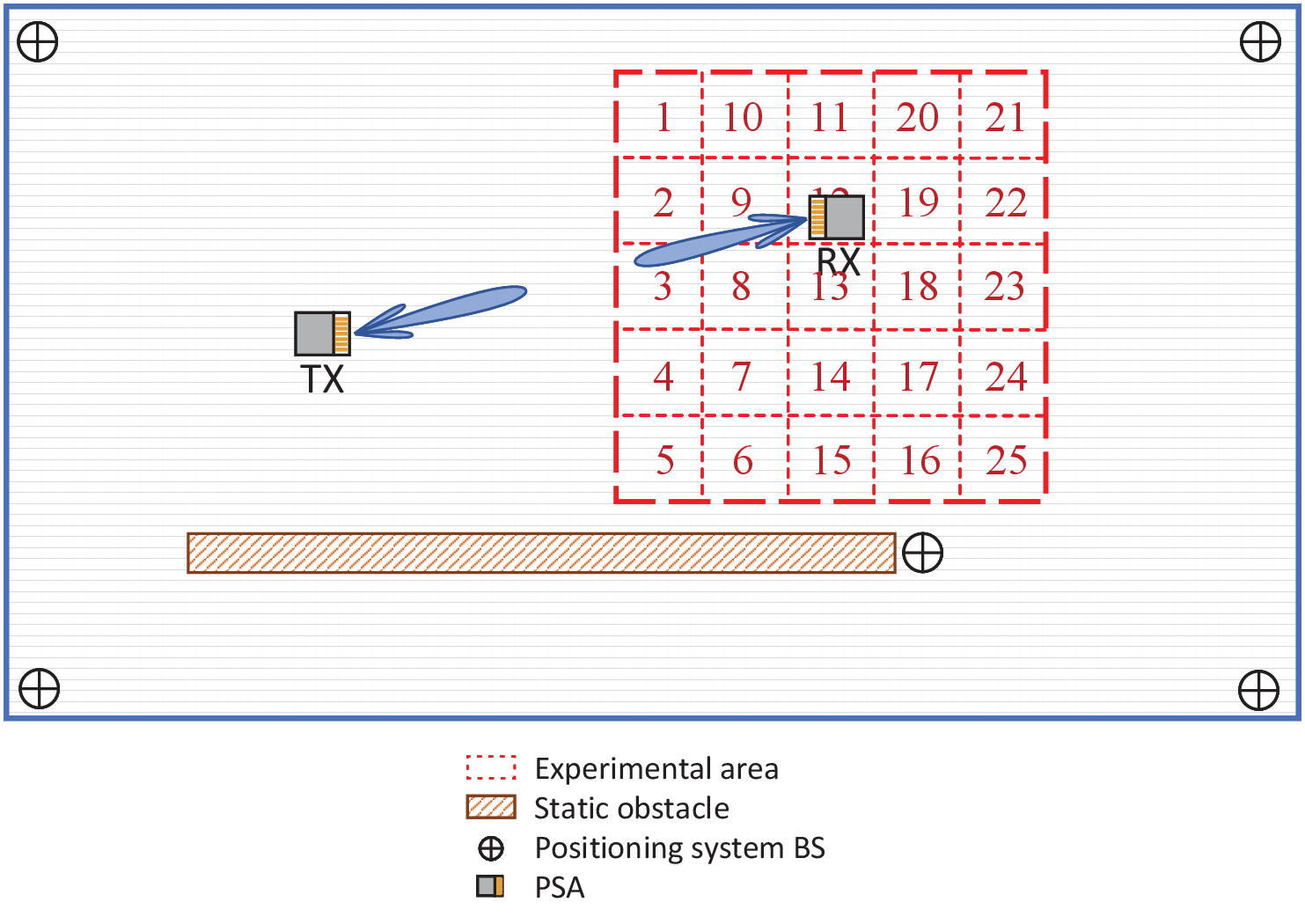}%
        \label{fig:1b}}
        \caption{CKM-based beam alignment with LoS link  in quasi-static environment.}
        \label{fig:1}
\end{figure}

With the   location-based beam alignment strategy, 
the corresponding prototype  was built in a
$4 \mathrm{m} \times 4 \mathrm{m} $ area. As shown in Fig. \ref{fig:1}, two 16-antenna 
mmPSAs are used as receiver and transmitter, respectively.
In the experimental area, there are five UWB base stations 
for positioning. UWB positioning tags are placed on the mmPSAs for real-time 
positioning of the transmitter and receiver.
With the   location information $\hat{\mathbf{b}}[t]=
[\hat{b}_{x}[t],\hat{b}_{y}[t]]$ and 
$\hat{\mathbf{q}}[t]=[\hat{q}_x[t],\hat{q}_y[t]]$ obtained from the UWB module,
 the angle    $ \phi[t] $ will be calculated in real time.
Further, with the  receiver orientation $\omega[t]$ with respect to the x-axis 
acquired by the gyroscope on the receiver, the location-based strategy 
will refine the beam index at the receiver to obtain better beam alignment.

For    mmPSAs at both the receiver and transmitter, the
beam index and pointing angle mapping codebook   has the following form as 
$ \big\{ m , \alpha(m )   \big\},m=1,...,M$, where $\alpha(m )$ is the beam angle,
denoting the angle between the beam center and the orientation of mmPSA,
and $m $ is the corresponding beam index.
By calculating the angle $ \phi[t] $ between the transmitter and receiver 
and obtaining the receiver orientation $\omega[t]$, the appropriate beam indexes can be found from 
the codebook.
In this case,  the transmit and receive beam indexes  can be respectively selected as
\begin{equation}\label{eq:istar}
        i^{\star}=\arg\min\limits_{i} \Big( \big| \phi[t]-\alpha(i) \big| \Big),
\end{equation}
\begin{equation}\label{eq:jstar}
        j^{\star}=\arg\min\limits_{j} \Big( \big| \phi[t]+\omega[t]-\alpha(j) \big| \Big),
\end{equation}
\begin{equation}\label{eq:hatft}
        \hat{\mathbf{f}}[t] =h\big( {i^{\star}}\big),
\end{equation}
\begin{equation}\label{eq:hatwt}
         \hat{\mathbf{w}}[t]=h\big( {j^{\star}}\big),
\end{equation}
where $h$ represents the conversion from  beam index to pre-set beamforming vector in
mmPSA. The detailed process of the location-based beam alignment strategy  
is summarized in Algorithm 1.

\begin{algorithm}[t]
        \caption{ Process of the location-based beam alignment strategy   }
        \label{alg:algorithm-1}
        \hspace*{0.02in}{\textbf{Input:  }}The beam codebook $\{ m , \alpha(m ) \big\}$ of mmPSA, the angle $ \phi[t] $ and the receiver orientation $\omega[t]$.\\
        \hspace*{0.02in}{\textbf{Output:  }} The transmit and receive beam
        indexes $i^{\star}$ and $j^{\star}$.
        \begin{algorithmic}[1]
            \State Initialization: Read the
            beam and pointing angle mapping codebook   
            $ \big\{ m , \alpha(m )   \big\}$.
            \For  {each    time slot   }
            \State Obtain the  location information of transmitter and receiver $\hat{\mathbf{b}}[t]=
            [\hat{b}_{x}[t],\hat{b}_{y}[t]]$ and 
            $\hat{\mathbf{q}}[t]=[\hat{q}_x[t],\hat{q}_y[t]]$ from the UWB module.
            \State Calculate the angle $\phi[t]$ between transmitter and receiver with (\ref{eq:phit}).
            \State Obtain the receiver orientation $\omega[t]$ from the gyroscope.
            \State Calculate the transmit and receive beam
            indexes $i^{\star}$ and $j^{\star}$ with  (\ref{eq:istar}) and (\ref{eq:jstar}).
            \EndFor  
        \end{algorithmic}
\end{algorithm}

In order to avoid repeatedly sending the same control commands, 
we add a beam index detection process to the beam alignment scheme.
Note that the selected beamforming vector $\hat{\mathbf{f}}[t]$ and $\hat{\mathbf{w}}[t]$ 
vary depending on the change in the position of the transmitter and receiver, not the time $t$.
For example, if $\hat{\mathbf{f}}[t+1]=\hat{\mathbf{f}}[t]$, the beam index 
does not change and there is no need to send a PSA control command to the transmitter; 
if $\hat{\mathbf{f}}[t+1]\neq \hat{\mathbf{f}}[t]$, 
then the control frame should be sent to the transmitter to modify the beam index.
The same applies to the receiver and its beamforming vector  $\hat{\mathbf{w}}[t]$.

Fig. \ref{fig:1a} and  Fig. \ref{fig:1b} show the actual scenario and the illustrative diagram  
of the location-based beam alignment strategy, respectively.
Although the location-based strategy 
is effective for LoS scenario depicted in
Fig. \ref{fig:1a}, it still needs to calculate 
the angle $ \phi[t] $  between the receiver and transmitter in real time.
Meanwhile, it is difficult for the  location-based  strategy  to handle environments with 
blocked LoS link or  dynamic environments.
The above mentioned problems will be solved accordingly in the CKM-based beam alignment strategy.

\subsection{CKM-based Beam Alignment in Quasi-Static Environment }

\begin{figure}[!t]
        \centering
        \subfloat[]{\includegraphics[width=1\columnwidth]{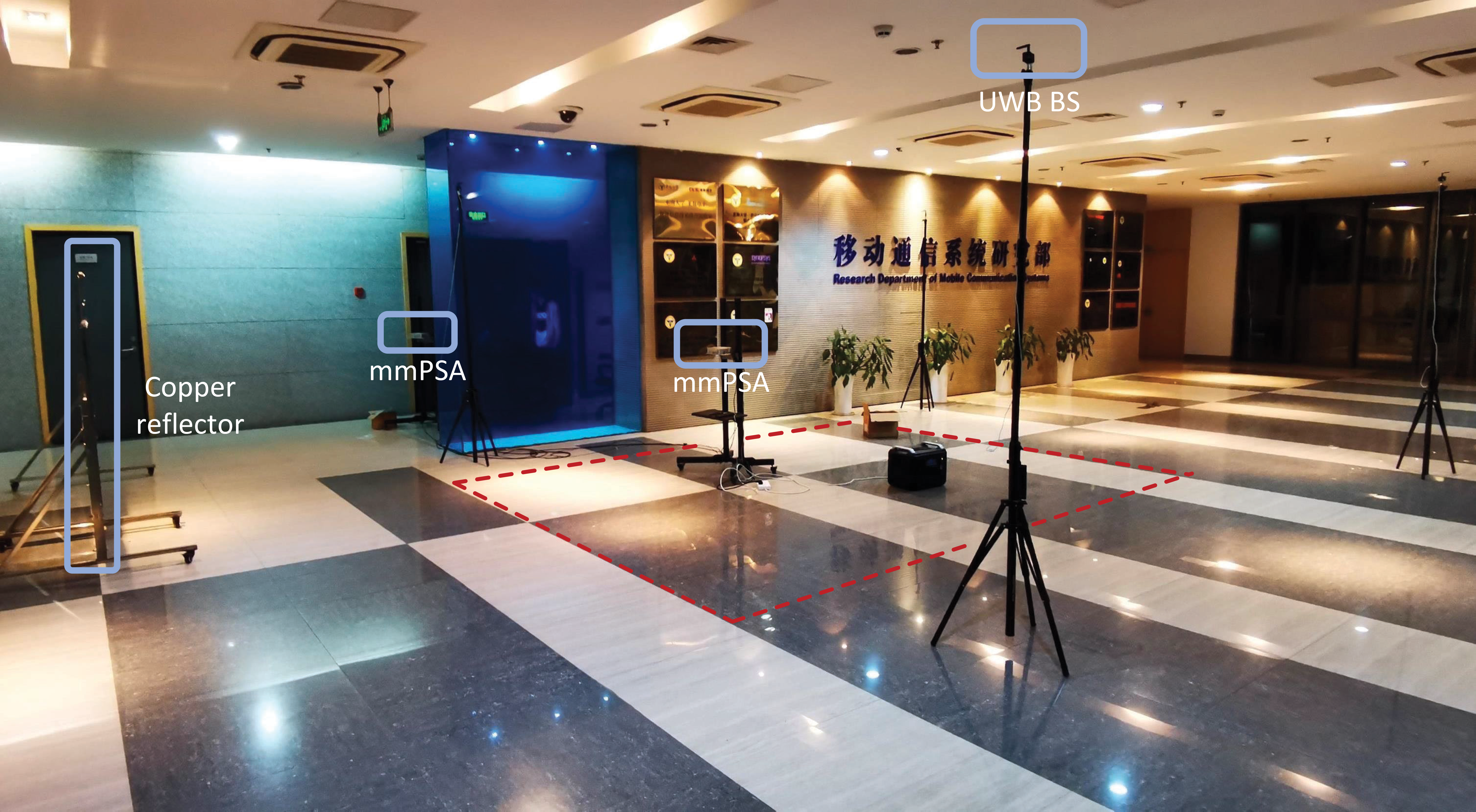}%
        \label{fig:2a}}
        \hfil
        \subfloat[]{\includegraphics[width=1\columnwidth]{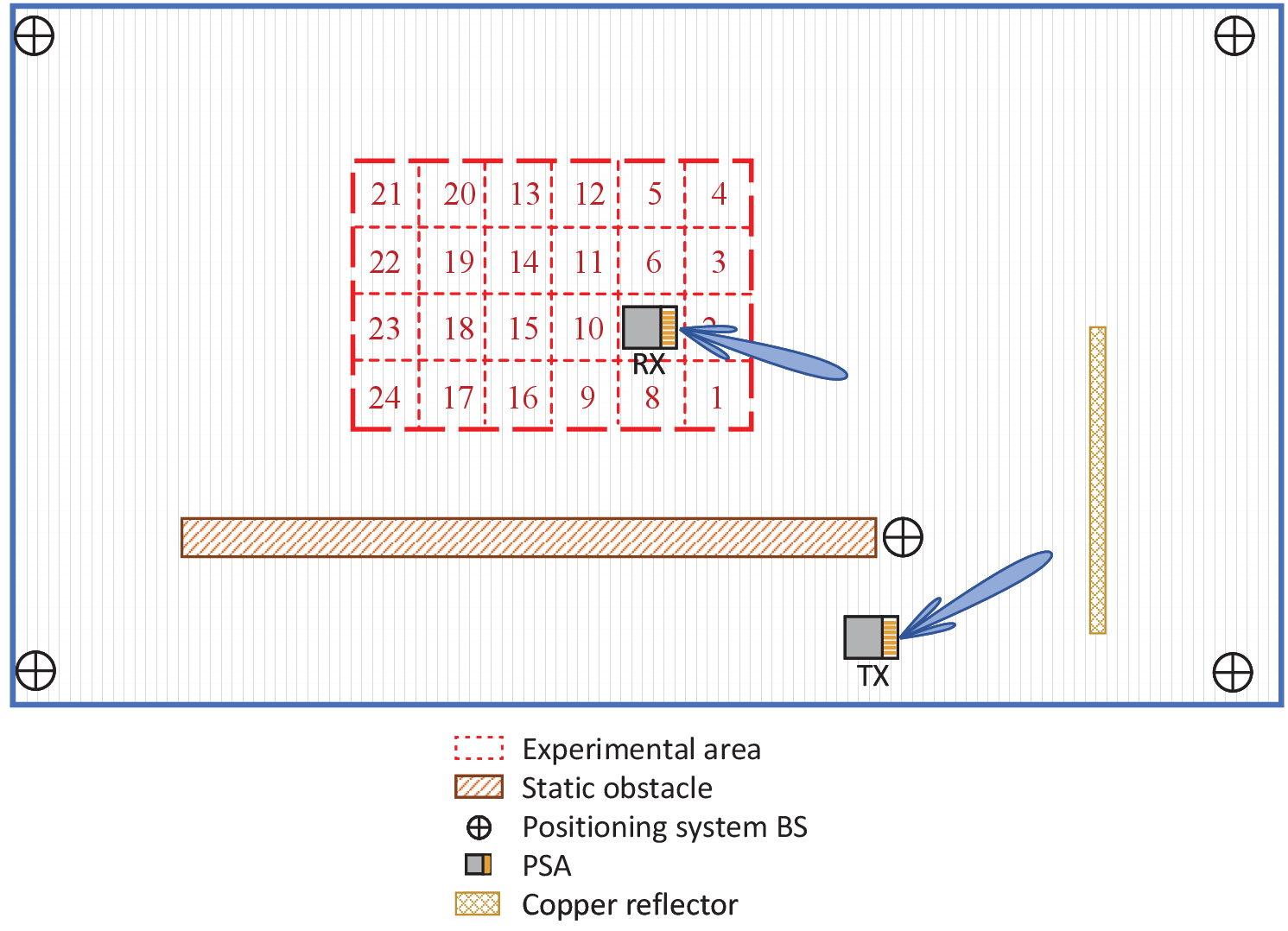}%
        \label{fig:2b}}
        \caption{ CKM-based beam alignment with NLoS link  in quasi-static environment.} 
        \label{fig:2}
\end{figure}

In this section, a comparison will be made between 
the CKM-based beam alignment strategy and the location-based benchmark strategy 
in quasi-static environments.
A typical LoS scenario and  a typical NLoS scenario are selected, respectively.
The corresponding CKMs are then constructed based on different environment information 
to help achieve beam alignment.
The LoS scenario is  identical to Fig. \ref{fig:1b}, 
with the receiver orientation $ \omega[t]$   changed arbitrarily.
The NLoS scenario is shown in Fig. \ref{fig:2}, where the orientation of  the receiver 
is fixed with $ \omega[t]=0$ for simplicity. In Fig. \ref{fig:2}, a concrete wall
with a thickness of $0.4m$  exists between the transmitter 
and the receiver to block the LoS link.
A copper reflector is placed  to create reliable 
reflective   link  
that can be utilized for communication.

\subsubsection{The construction of CKM}
In order to use environment  information to help  beam alignment,
we fix the transmitter location and build a CKM  
for the area of interest for the receiver. 
In Fig. \ref{fig:1b}, the CKM is 
built on a square area of size $4m \times 4m$, 
while in Fig. \ref{fig:2}, the CKM is 
built on a square area of size $3.2m \times 4.8m$. 
The grid size is $0.8m\times 0.8m$.
For each   grid in CKM, the center coordinate is used as the grid's coordinate.
In the offline construction of the CKM, the exhaustive beam sweeping method 
is used to determine the optimal beam index pairs. 
During the construction
of CKMs, the receiver orientation is set to  $ \omega[t]=0$.
Considering that the construction steps of CKM are universal to LoS and NLoS scenarios, 
we take the NLoS scenario as an example and the specific process   is as follows:
First, the position of the transmitter is fixed, and the grids within the 
entire CKM are numbered   from 1 to 24 in Fig. \ref{fig:2b}.
Then, the UWB tag is placed on the receiver, 
and the receiver is sequentially moved  from    grid 1 to   grid 24
in sequence to find the optimal transmit and receive beam index pairs 
for  the grid.

When the receiver moves to the center of grid   $i$, 
the coordinate $\big(x_{i},y_{i} \big)$  
of the receiver is obtained  by the UWB system  and stored in the CKM. After that, 
the beam index pair of the transmitter and receiver is switched
to perform exhaustive beam sweeping. The difference between the diversified 
beam index pairs is reflected in the power of the signal received at the receiver. 
The higher the power of the received signal, the better the corresponding beam index pair. 
For each  grid,   communication tests with 64x64 beam index pairs are conducted, 
since each mmPSA has 64 preset beams in the codebook.
The communication test for each beam pair lasts 30$\mathrm{ms}$, 
during which the received signal power is read three times in 10$\mathrm{ms}$ intervals 
and the maximum received signal power is taken to reduce the impact of 
beam index pair switching on the communication performance.
The transmit and receive beam index pair  with the maximum received power 
will be selected and stored in  the CKM as the optimal transmit and receive beam index 
pair $\big(m^{\mathrm{t}}_{i},
m^{\mathrm{r}}_{i} \big)$ of the reflective NLoS link  for    grid $i$. Therefore, the corresponding CKM 
for   grid  $i$ is recorded 
and saved in the following form as
\begin{equation}
        c_{i}=\Big\{ \big(x_{i},y_{i} \big) ,\big(m^{\mathrm{t}}_{i},
        m^{\mathrm{r}}_{i} \big) \Big\}.
\end{equation}

The complete CKM construction process of the CKM-based beam alignment  
is summarized in Algorithm 2.
The construction step of the CKM does not change even in complex wireless environments, 
i.e., sweeping the transmit and receive beam pairs in the codebook. 
Therefore, when the number $s_g$  of grids on the sides of a square area increases,
which means the size of the environment increases, the construction complexity of the 
CKM is only $\mathcal{O}(s_c s_{g}^{2}) $, where  $s_c$ is the size of the codebook.
This suggests that the complexity of construction does not depend on the complexity 
of the wireless environment.

\begin{algorithm}[t]
        \caption{ General CKM construction process   }
        \label{alg:algorithm-2}
        \hspace*{0.02in}{\textbf{Input:  }}The beam codebook $\{ m , \alpha(m )   \big\}$ of mmPSA.\\
        \hspace*{0.02in}{\textbf{Output:  }} The corresponding CKM   $\{c_{i} \}$ for  the grid.
        \begin{algorithmic}[1]
            \State Initialization: Fix the position of the transmitter, number the 
             grids.
            \For  {each    grid $i$  }
            \State Move the receiver to the center of  grid $i$.
            \State Obtain the receiver coordinate $\big(x_{i},y_{i} \big)$ by UWB.
            \For  {each transmit beam index}
            \For {each receive beam index}
            \State Conduct a   communication test, record the 
            received signal power  at  regular intervals,
            \State Select the maximum received power from the three records 
            for each beam index pair.
            \EndFor
            \EndFor
            \State Select the optimal beam index pair $\big(m^{\mathrm{t}}_{i},
            m^{\mathrm{r}}_{i} \big)$ with the maximum received power.
            \State Save the   CKM for    grid $i$   as 
            $c_{i}=\Big\{ \big(x_{i},y_{i} \big) ,\big(m^{\mathrm{t}}_{i},
            m^{\mathrm{r}}_{i} \big) \Big\}$.
            \EndFor  
        \end{algorithmic}
\end{algorithm}

\subsubsection{CKM experiment with LoS link}

\begin{figure}
        \centering
         \includegraphics[width=1\columnwidth]{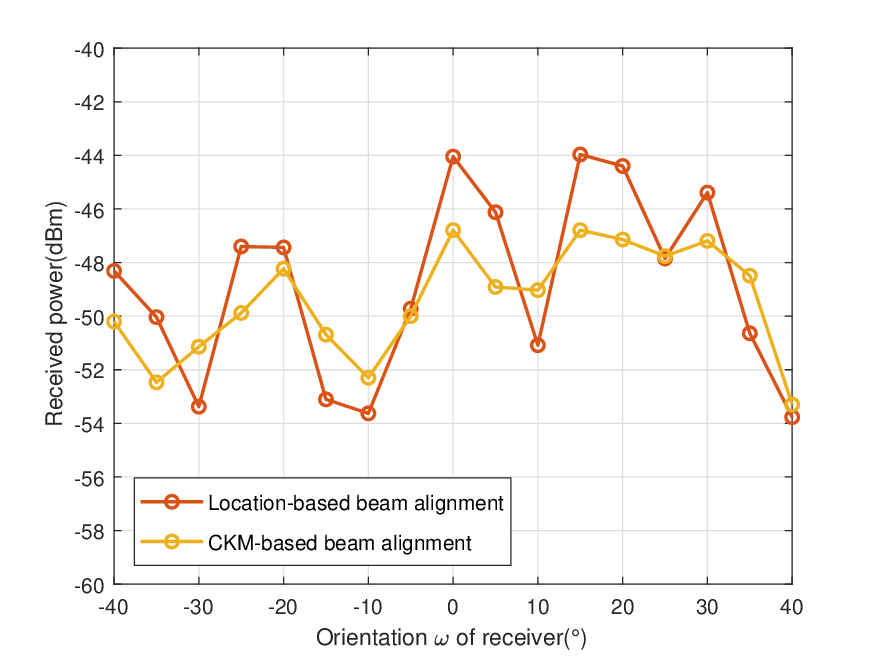}
        \caption{
                Performance comparison of CKM-based and Location-based strategies 
        under different receiver orientations.}
        \label{fig:exp1result}
      \end{figure}

In the static scenario with  LoS link,   the experiment  with 
an arbitrary   grid $i$ in the CKM is conducted. 
The coordinates  of the receiver are obtained by the UWB positioning system,
while the orientation $\omega[t]$ of the receiver is obtained by a gyroscope fixed on the receiver.
The geometric representation of the CKM-based strategy in LoS scenario is also shown in 
Fig. \ref{fig:location}.
With continuously changing receiver orientation, 
 the CKM-based beam alignment strategy can predict the transmit and 
receive beam pairs based on channel knowledge and receiver orientation.
Specifically, the coordinate $\hat{\mathbf{q}}[t]=[\hat{q}_x[t],\hat{q}_y[t]]$
of the receiver can be obtained in real time through the UWB positioning system.
By calculating the   distance between the receiver and the center of 
each grid,  we can find out which grid center 
is nearest to the real-time position of the receiver 
and thus determine which grid $i^{\star}$ the receiver 
is located in real time as
\begin{equation}\label{eq:blocki}
        i^{\star}=\arg\min \limits_{i} \Big(  \big(\hat{q}_x[t]-x_{i}\big)^{2}
        +\big(\hat{q}_y[t]-y_{i}\big)^{2}  \Big).
\end{equation}

Then, the transmit beam index $m^{\mathrm{t}}_{i^{\star}}$ of grid $i^{\star}$ is obtained according to 
the constructed CKM, and the receive beam index is refined according to 
the receiver orientation $\omega[t]$ as 
\begin{equation}\label{eq:mri'}
        m^{\mathrm{r}}_{i^{'}}=\arg\min\limits_{j} \Big( \big|   
        \alpha(m^{\mathrm{r}}_{i^{\star}})+\omega[t]-
        \alpha(j) \big| \Big).
\end{equation}

Fig. \ref{fig:exp1result} shows the performance comparison of the two 
beam alignment strategies for different receiver orientations. 
The receiver orientation  $ \omega[t] $ ranges from -$40^{\circ}$ to  $40^{\circ}$ at the step of $5^{\circ}$.
As can be seen from Fig. \ref{fig:exp1result}, 
for different receiver orientations, the CKM-based beam alignment strategy refines 
the beam index pairs read from the CKM according to orientation $\omega[t]$, 
and the received power obtained is very close to that of the location-based strategy.
Therefore,
the CKM-based beam alignment strategy 
is able to maintain good communication performance  regardless of the orientation of the receiver.
This is because the CKM-based strategy exploits the channel knowledge
to first obtain the original transmit and receive beam index pairs stored in the CKM.
Then the beam index pair is refined according to the receiver orientation
to achieve good beam alignment.
Therefore, in quasi-static scenarios with LoS links, the CKM-based beam alignment 
strategy can obtain comparable communication performance to the 
location-based benchmark strategy.
The performance difference between the two strategies mainly comes from the 
location  error and the gyroscope inertia error
in different receiver orientations. 
The CKM-based strategy has smaller variation mainly because  the
location errors have a large impact on the location-based strategy, 
but generally do not interfere with the judgment of the nearest grid center, 
and therefore interfere less with the CKM-based strategy.

\subsubsection{CKM experiment with NLoS link}


The relevant information of the communication environment is embedded in the CKM  $\{c_{i} \}$
and is expressed by the optimal beam index pair for each grid.
During the experiment, the receiver can move freely within the constructed  CKM area, 
while the prototype performs beam alignment through the already constructed CKM.
For convenience, the receiver orientation is set to $\omega[t]=0$.
It is worth mentioning that the receiver orientation $\omega[t]=0$ 
is assumed in order to demonstrate the beam alignment performance of the 
CKM-based strategy through NLoS links in different grids under uniform conditions.
When $\omega[t]\neq 0$, the receive beam index read from the CKM should be refined 
according to  $\omega[t]$, which is consistent with (\ref{eq:mri'}) in the LoS scenario.
Specifically, the grid $i^{\star}$ to which the receiver belongs is first determined 
by the Euclidean distance of the receiver from each grid
with (\ref{eq:blocki}).
Similarly, the transmit beam index $m^{\mathrm{t}}_{i^{\star}}$ is obtained from the 
constructed CKM. 
Considering $\omega[t]=0$ in (\ref{eq:mri'}), the index 
$m^{\mathrm{r}}_{i^{\star}}$ of grid $i^{\star}$ can   be obtained
directly from the CKM as the receiver beam index.
A similar beam index detection process is added to the CKM-based  
beam alignment, i.e., the control frames are retransmitted only when the 
transmit or receive beam index changes.

\begin{figure}
        \centering
         \includegraphics[width=1\columnwidth]{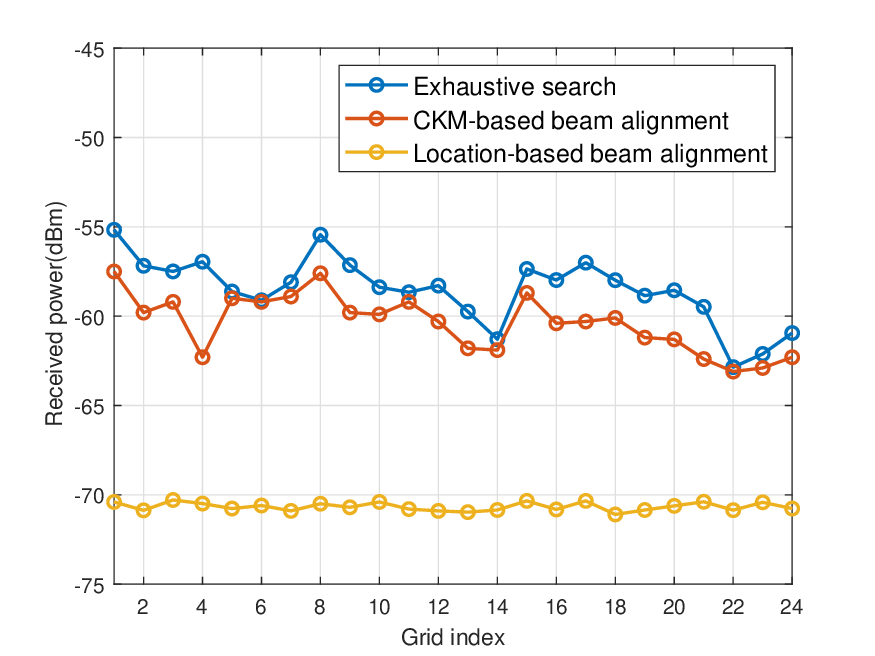}
        \caption{Performance comparison between CKM-based and location-based strategies in NLoS scenario.}
        \label{fig:exp2aresult}
      \end{figure}

The  actual scenario and the illustrative diagram  of the CKM-based
beam alignment prototype are respectively shown in Fig. \ref{fig:2a} and Fig. \ref{fig:2b}.
In Fig. \ref{fig:exp2aresult}, the communication performance of CKM-based beam alignment
is compared with the benchmark location-based  and   beam sweeping strategies.
As illustrated in  Fig. \ref{fig:exp2aresult}, the location-based strategy
gives poor performance because it can only assume the LoS 
links that do not exist in the considered scenario. Meanwhile,
since the reflective link considering the environment information is stored 
in the CKM in the form of beam index pairs, the
CKM-based beam alignment strategy  can use the reflective link  for communication 
even if the LoS link is blocked.
By comparing the received power at each grid, 
the CKM-based strategy is able to achieve a communication performance 
comparable to that of the exhaustive beam sweeping strategy, but with almost 
no real-time overhead.
The modest performance loss is caused primarily by the receiver's location error 
relative to the grid center during movement.

\subsection{CKM-based Beam Alignment in Dynamic Environment}

\begin{figure}[!t]
    \centering
    \subfloat[]{\includegraphics[width=1\columnwidth]{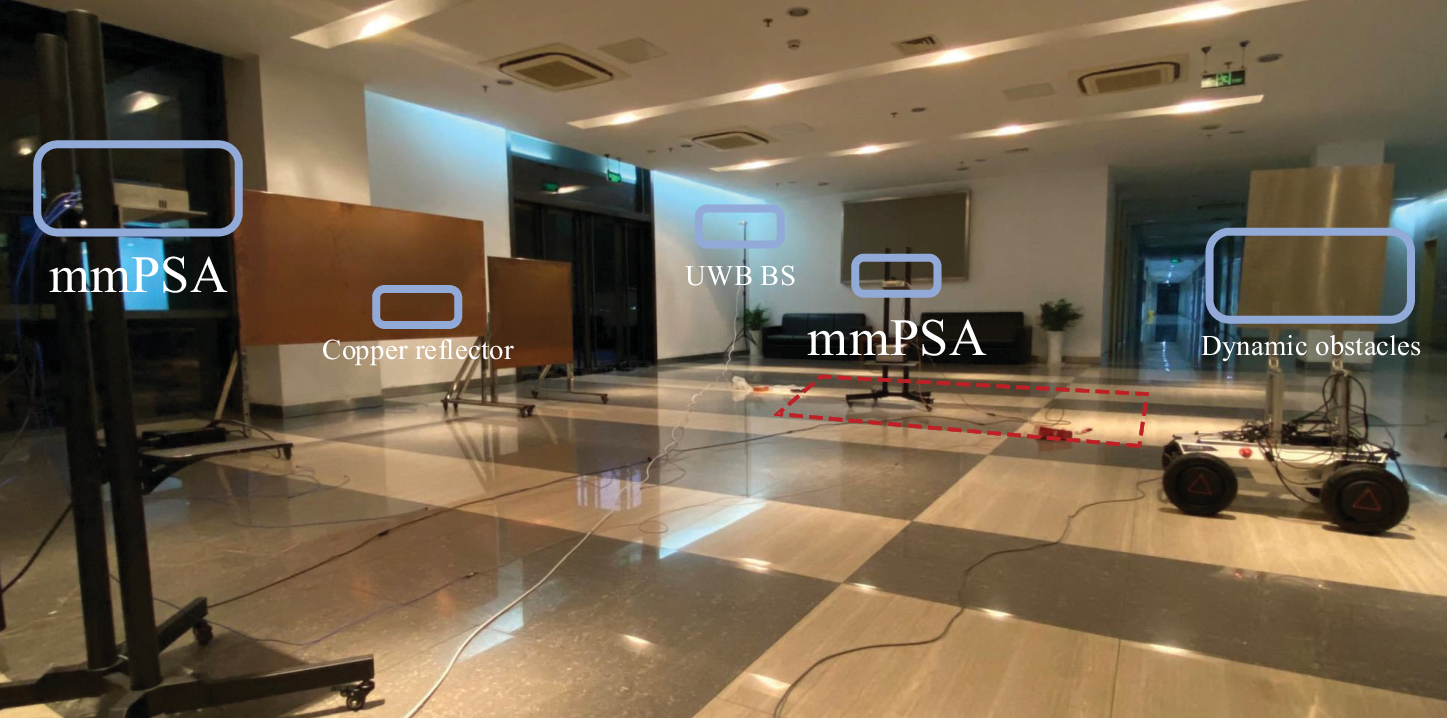}%
    \label{fig:3a}}
    \hfil
    \subfloat[]{\includegraphics[width=1\columnwidth]{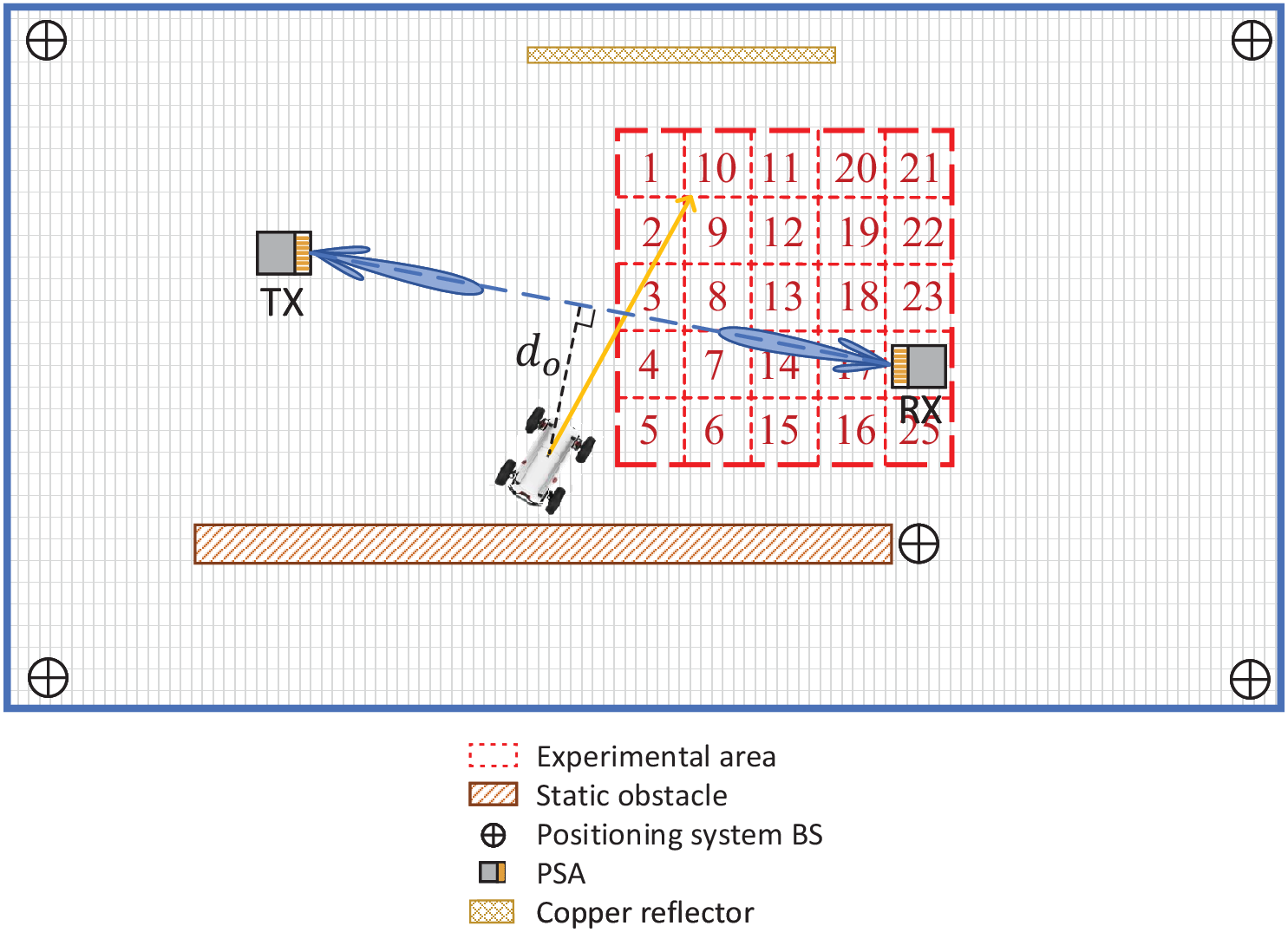}%
    \label{fig:3b}}
    \hfil
    \subfloat[]{\includegraphics[width=1\columnwidth]{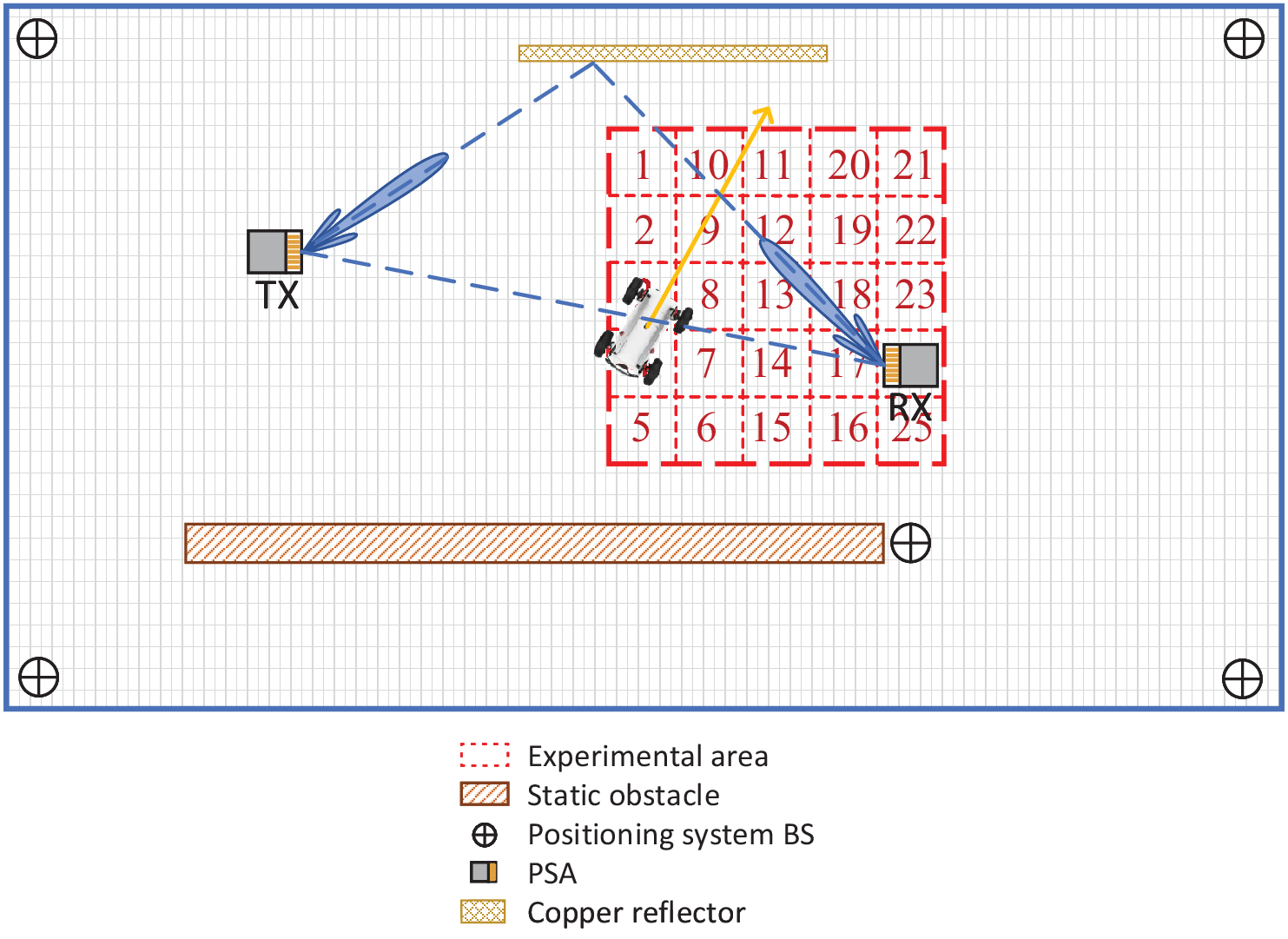}%
    \label{fig:3c}}
    \caption{CKM-based beam alignment    in dynamic environment}
    \label{fig:3}
\end{figure}

CKM-based beam alignment is effective not only in quasi-static environments, 
but also in dynamic ones.
In this experiment, a typical dynamic environment is considered, 
and the prototype's CKM-based beam alignment strategy is enhanced to 
increase the applicability of environment-aware communication.
The experimental scenario  
is shown in Fig. \ref{fig:3a}. The   receiver maintains the fixed orientation
with   $ \omega[t]=0$. Meanwhile, a copper reflector  
is fixed on the side to simulate an  NLoS  link in a complex environment. 
Therefore, in this experimental scenario, both LoS and NLoS links may 
exist simultaneously.
The communication link must be determined in real time 
based on the current condition of the dynamic environment. 
The UWB positioning system locates targets in the scenario by UWB tags.
A $50\mathrm{cm}\times 70\mathrm{cm}$ small aluminum reflective plate is fixed 
on the moving vehicle to simulate the dynamic obstacles in the scenario.
The moving vehicle is fitted with an auxiliary UWB tag to simulate dynamic obstacles.

\subsubsection{The construction of CKM}
The CKM in this experiment is built on a $4m \times 4m$ square area 
with $0.8m\times 0.8m $ grid. 
The process of  CKM construction in a dynamic environment is similar to 
Algorithm 1.
The main   difference in the construction of CKM from  Algorithm 1 is that, 
only the reflective NLoS link in the environment is considered, i.e., 
the optimal 
beam index pair $\big(m^{\mathrm{t}}_{i},
m^{\mathrm{r}}_{i} \big)$ needs to be found  and stored  in CKM. 
However, in the construction of CKM in the dynamic environment, both LoS and NLoS links 
need to be considered in order to improve the ability of the prototype to 
maintain communication in dynamic environments.
Therefore, in the exhaustive search at each grid $i$,  
the optimal LoS link beam index pair $\big(m^{\mathrm{t}}_{i},
m^{\mathrm{r}}_{i} \big)$ and reflective NLoS link beam index pair  $\big(l^{\mathrm{t}}_{i},
l^{\mathrm{r}}_{i} \big)$ need to be found out
in the spatial direction according to the recorded received signal power and codebook, 
and stored in the corresponding CKM with the following format as 
\begin{equation}
    c_{i}=\Big\{ \big(x_{i},y_{i} \big) ,\big(m^{\mathrm{t}}_{i},
    m^{\mathrm{r}}_{i} \big) ,\big(l^{\mathrm{t}}_{i},
    l^{\mathrm{r}}_{i} \big)\Big\}.
\end{equation}

\subsubsection{CKM experimental procedures and results}

The simulation of the dynamic environment and the process of CKM-based beam alignment 
in the corresponding dynamic environment are presented here. 
During the experiment, the moving vehicle with an aluminum reflective plate 
will serve as a significant communication environment-blocking dynamic obstacle.
The receiver can communicate with the transmitter at any location within the 
constructed CKM area, and during the communication, we will move the moving vehicle 
to introduce dynamic blockage.
The moving vehicle first moves toward the LoS link between  transmitter and receiver, 
as in Fig. \ref{fig:3b}.
Considering that the LoS link  is currently unimpeded, 
this phase is used to simulate a scenario where the dynamic environment does not affect 
communication temporarily.
Next, we control the moving vehicle carrying an aluminum reflective plate
across the LoS between the transmitter and receiver. As shown in Fig. \ref{fig:3c}, 
the moving vehicle remains stationary for a period of time as it 
passes through the LoS link to simulate a scenario in which the LoS link is blocked. 
After that, the vehicle will continue to move until it is far away from the LoS link, 
which is used to represent the moment when the dynamic environment 
changes again and the LoS link recovers.

After obtaining the grid index, the key to CKM-based beam alignment 
in dynamic environments is determining which beam index pair in the CKM will 
be used for communication.
Denote the real-time coordinate of the dynamic obstacle  as 
$\hat{\mathbf{o}}[t]=[\hat{o}_x[t],\hat{o}_y[t]]$. In order to 
avoid the communication outage due to  the dynamic obstacle, 
we use the distance-based judging rule to select the communication link.
Specifically, the LoS link is used by default 
in the communication. Once the distance from the dynamic obstacle to the LoS link 
is less than a given threshold $\eta$, the selected communication
link  will 
switch from the LoS link to the reflective NLoS link in advance 
to avoid   communication outage.
To represent the position relation between dynamic obstacle and transmitter and receiver, 
we introduce the following vector according to the coordinates obtained from UWB positioning system
\begin{equation}
        \begin{aligned}
                \overrightarrow{\mathrm{TR}}&=\hat{\mathbf{q}}[t]-\hat{\mathbf{b}}[t]
                \\ &=\Big[\hat{q}_x[t]-\hat{b}_x[t],\hat{q}_y[t]-\hat{q}_y[t]\Big],
        \end{aligned}
\end{equation}
\begin{equation}
        \begin{aligned}
                \overrightarrow{\mathrm{TO}}&=\hat{\mathbf{o}}[t]-\hat{\mathbf{b}}[t]
                \\ &=\Big[\hat{o}_x[t]-\hat{b}_x[t],\hat{o}_y[t]-\hat{q}_y[t]\Big],
        \end{aligned}
\end{equation}
where  $\hat{\mathbf{b}}[t]=[\hat{b}_x[t],\hat{b}_y[t]]$,$\hat{\mathbf{q}}[t]=[\hat{q}_x[t],\hat{q}_y[t]]$
and $\hat{\mathbf{o}}[t]=[\hat{o}_x[t],\hat{o}_y[t]]$ denote the real-time
coordinates  of the transmitter, receiver and dynamic obstacle, respectively.
$ \overrightarrow{\mathrm{TR}}$ represents the vector pointing from the 
transmitter to the receiver, while $\overrightarrow{\mathrm{TO}}$ represents 
the vector pointing from the transmitter to the dynamic barrier.

\begin{figure}
        \centering
         \includegraphics[width=1\columnwidth]{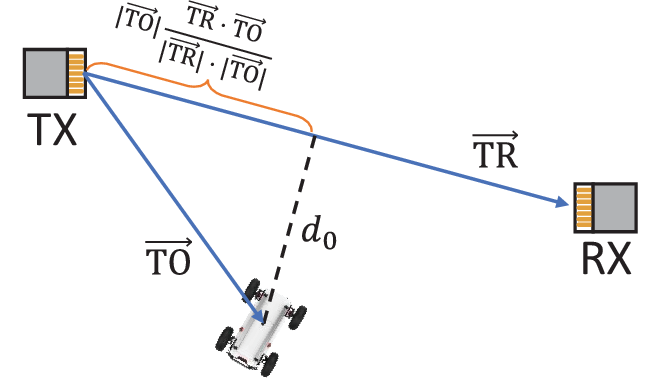}
        \caption{A geometric illustration of the relationship between dynamic obstacle and devices.}
        \label{fig:vector}
      \end{figure}

As shown in Fig. \ref{fig:vector},  the ratio $r$  of the length of the projection of $\overrightarrow{\mathrm{TO}}$  on $\overrightarrow{\mathrm{TR}}$
to the length of $\overrightarrow{\mathrm{TR}}$ is calculated in (\ref{eq:ratio}), while  the distance $d_{\mathrm{o}}$ 
from the dynamic obstacle to the LoS link is expressed as in (\ref{eq:distanceo})
\begin{equation}\label{eq:ratio}
        r=\frac{|\overrightarrow{\mathrm{TO}}| \frac{\overrightarrow{\mathrm{TR}} \cdot
        \overrightarrow{\mathrm{TO}}}{|\overrightarrow{\mathrm{TR}}|\cdot |\overrightarrow{\mathrm{TO}}|} }
        {|\overrightarrow{\mathrm{TR}}  |},
\end{equation}
\begin{equation}\label{eq:distanceo}
        |d_{\mathrm{o}}|=\big| \overrightarrow{\mathrm{TO}}-r\overrightarrow{\mathrm{TR}} \big|.
\end{equation}

A complete CKM-based beam alignment process in dynamic environments is shown in Algorithm 3.
\begin{algorithm}[t]
        \caption{ CKM-based beam alignment process in dynamic environment  }
        \label{alg:algorithm-3}
        \hspace*{0.02in}{\textbf{Input:  }} The constructed CKM $\{c_{i}\}$ and the distance threshold $\eta$. \\
        \hspace*{0.02in}{\textbf{Output:  }} The beam index pair  $\big(m^{\mathrm{t}}[n],
        m^{\mathrm{r}}[n]\big)$.
        \begin{algorithmic}[1]
            \State Initialization: obtain the  real-time coordinates 
        $\hat{\mathbf{b}}[t]$,$\hat{\mathbf{q}}[t]$
        and $\hat{\mathbf{o}}[t]$ from UWB.
            \State Determine the grid $i^{\star}$ in  CKM 
            to which the receiver belongs by (\ref{eq:blocki}).
            \State Calculate the ratio $r$ with (\ref{eq:ratio}).
            \State Calculate the distance $d_{\mathrm{o}}$ 
            from the dynamic obstacle to the LoS link with (\ref{eq:distanceo}).
            \If  {$0<r<1$ and $|d_{\mathrm{o}}|\leq \eta$,}
            \State  $
                \hat{\mathbf{f}}[t]=h\big(m^{\mathrm{t}}_{i^{\star}}\big),$
            \State  $ \hat{\mathbf{w}}[t]=h\big(m^{\mathrm{r}}_{i^{\star}}\big),$
            \Else
            \State  $
                \hat{\mathbf{f}}[t]=h\big(l^{\mathrm{t}}_{i^{\star}}\big),$
            \State  $ \hat{\mathbf{w}}[t]=h\big(l^{\mathrm{r}}_{i^{\star}}\big).$
            \EndIf 
        \end{algorithmic}
\end{algorithm}

\begin{figure}
        \centering
         \includegraphics[width=1\columnwidth]{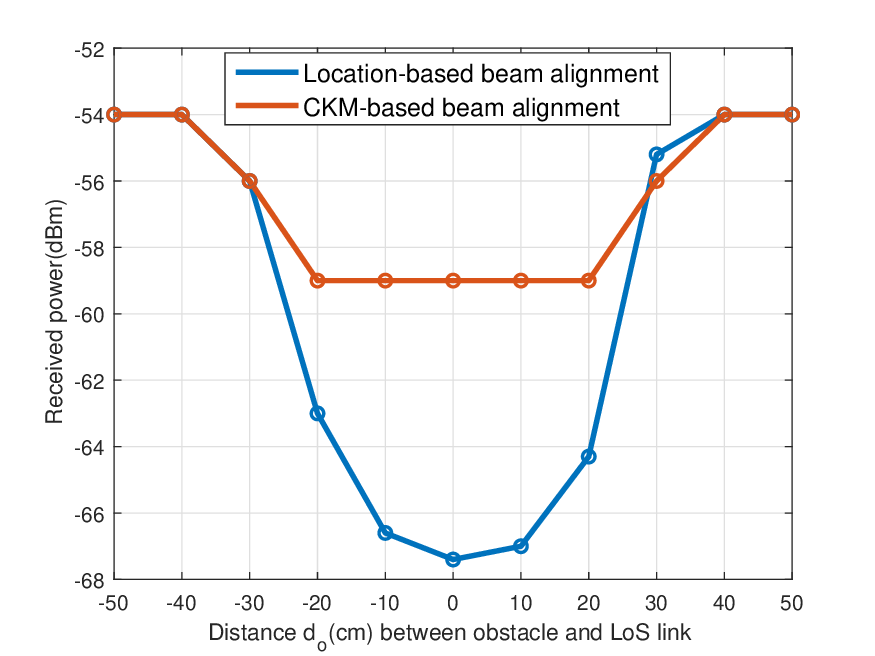}
        \caption{Performance comparison between CKM-based and location-based strategies 
        in dynamic environment.}
        \label{fig:environmentalarawareness}
      \end{figure}

In Fig. \ref{fig:environmentalarawareness}, we compare the performance of the 
 CKM-based beam alignment with the benchmark location-based beam alignment in the dynamic environment.
The received power is recorded versus different  distances $d_{\mathrm{o}}$  from the dynamic obstacle to the LoS link
with the step $10$cm. 
Considering that half of the width of the aluminum reflective plate on the 
moving vehicle is $25$cm, the distance threshold is set as $\eta=30$cm. 
When the moving vehicle representing the dynamic obstacle just 
starts to move, it is far away from the LoS link, which is shown in Fig. \ref{fig:environmentalarawareness}
with $-50\mathrm{cm}\leq d_{\mathrm{o}}<-30\mathrm{cm}$. In this case, 
the LoS link is almost undisturbed by the 
dynamic obstacle, and  the CKM-based beam alignment strategy chooses the LoS link with
$\big(m^{\mathrm{t}}_{i}, m^{\mathrm{r}}_{i} \big) $ for communication,
 which is consistent with the results 
 obtained by the location-based strategy. This explains the coincidence of the two received power 
lines with $-50\mathrm{cm}\leq d_{\mathrm{o}}<-30\mathrm{cm}$. When $-30\mathrm{cm}\leq d_{\mathrm{o}} \leq 30\mathrm{cm}$, which means 
$| d_{\mathrm{o}}| < \eta$, different beam index pairs are selected for communication 
by the benchmark location-based  and the CKM-based strategies. On the one hand, the location-based strategy has 
no  knowledge of the dynamic environment and can only choose the beam index pair of the  LoS link. 
Thus, as the distance $d_{\mathrm{o}}$ changes from $-30$cm to $30$cm, the LoS link is first 
gradually obstructed by the dynamic obstacle and then gradually recovers, 
which also leads to a decrease and then an increase in the received power and a corresponding  
minimum at $d_{\mathrm{o}}=0$cm. On the other hand,
the CKM-based beam alignment strategy fully considers the impact of dynamic obstacle 
and switches to NLoS beam index pair $ \big(l^{\mathrm{t}}_{i}, l^{\mathrm{r}}_{i} \big)$ 
in time ($|d_{\mathrm{o}}|\leq \eta$). 
When the CKM-based strategy switches to the NLoS link for communication, 
the corresponding received power has a power loss up to 5dB, 
which corresponds to the  path loss between the alternative NLoS link and the LoS link.
Compared with the location-based strategy, 
the resulting gain of the received power is up to $8$dB or more.
This fully demonstrates that the CKM-based beam alignment can effectively cope with the 
variation of environment and improve communication performance 
through pre-stored channel knowledge.
Notice that the location-based strategy performs better at $d_{\mathrm{o}}=30$cm. 
This can be explained by the spatial limitations of the experimental site, 
where the dynamic obstacle moves closer to the copper reflector as the moving vehicle moves, 
thereby impacting the NLoS link and diminishing its performance to some degree.
As the distance increases to $d_{\mathrm{o}}\geq 30$cm, the CKM-based strategy 
treats the impact of dynamic obstacle on the LoS link as negligible and switches 
the beam index pair back to the LoS link 
$\big(m^{\mathrm{t}}_{i}, m^{\mathrm{r}}_{i} \big) $. Therefore, the curves  of the 
received power are once again consistent for both the benchmark and the CKM-based beam 
alignment.

\section{Conclusion}
In this paper, a prototype system  is developed for CKM-based environment-aware 
mmWave beam alignment. 
CKM makes full use of the   prior  information about the  actual environment 
to build a database that effectively reflects channel knowledge, 
thus enabling training-free environment-aware beam alignment. 
We first set up a general offline CKM construction method.
Then, by hardware integration and program design, a CKM-based mmWave beam alignment
 environment-aware communication prototype is built. 
The communication performance of the CKM-based beam alignment strategy is compared 
 with the location-based benchmark strategy under various quasi-static and dynamic scenarios, 
 effectively revealing the effectiveness and feasibility of the CKM-based mmWave 
 beam alignment system.


\bibliographystyle{IEEEtran}

\bibliography{Experiments2}

\vfill

\end{document}